\documentclass[multphys,vecphys]{svmult}

\usepackage{makeidx}     
\usepackage{graphicx}    
\usepackage{multicol}    

\def\asec{\ifmmode ^{\prime\prime}\else$^{\prime\prime}$\fi}

\def\msun{M$_{\odot}$}

\def\degs{\ifmmode ^{\circ}\else$^{\circ}$\fi}
\def\amin{\ifmmode ^{\prime}\else$^{\prime}$\fi}
\def\asec{\ifmmode ^{\prime\prime}\else$^{\prime\prime}$\fi}

\def\degs{\ifmmode ^{\circ}\else$^{\circ}$\fi}
\def\amin{\ifmmode ^{\prime}\else$^{\prime}$\fi}

\def\EE#1{\times 10^{#1}}

\def\cm3{\rm ~cm^{-3}}
\def\kms{\rm ~km~s^{-1}}

\def\wl{~\lambda}

\def\muJ{\mu{\rm Jy}}
\def\Msun{~{\rm M}_\odot}
\def\Msunyr{~{\rm M}_\odot~{\rm yr}^{-1}}
\def\Mdot{\dot M}
\def\EE#1{\times 10^{#1}}
\def\cm3{\rm ~cm^{-3}}
\def\kms{\rm ~km~s^{-1}}

\def\lsim{\!\!\!\phantom{\le}\smash{\buildrel{}\over
  {\lower2.5dd\hbox{$\buildrel{\lower2dd\hbox{$\displaystyle<$}}\over
                               \sim$}}}\,\,}
\def\gsim{\!\!\!\phantom{\ge}\smash{\buildrel{}\over
  {\lower2.5dd\hbox{$\buildrel{\lower2dd\hbox{$\displaystyle>$}}\over
                               \sim$}}}\,\,}

\makeindex             

\begin{document}

\title*{High-resolution radio imaging of young supernovae}
\author{M.A.~P\'erez-Torres\inst{1}
\and    J.M.~Marcaide\inst{2}
\and    A.~Alberdi\inst{1}
\and    E.~Ros\inst{3}
\and    J.C.~Guirado\inst{2}
\and    L.~Lara\inst{1,4}
\and    F.~Mantovani\inst{5}
\and    C.J.~Stockdale\inst{6}
\and    K.W.~Weiler\inst{6}
\and    P.J.~Diamond\inst{7}
\and    S.D.~Van Dyk\inst{8}
\and    P.~Lundqvist\inst{9}
\and    N.~Panagia\inst{10}
\and    I.I.~Shapiro\inst{11}
\and    R.~Sramek\inst{12}
}
\authorrunning{M.A.~P\'erez-Torres et al.} 

\institute{
IAA - CSIC, 
Apdo. Correos 3004, 
18008 Granada, Spain
\texttt{torres@iaa.es}
\and
Departamento de Astronom\'{\i}a, Universidad de Valencia, 46100 Burjassot,
Spain
\and
Max-Planck-Institut f\"ur Radioastronomie, 
53121 Bonn, Germany
\and
Dpto. de F\'{\i}sica Te\'orica y del Cosmos, 
Universidad de Granada, Spain
\and
Istituto di Radioastronomia/CNR, via P. Gobetti 101, 40129 Bologna,
Italy
\and
Naval Research Laboratory, Code 7213, Washington, DC 20375-5320, USA
\and
MERLIN/VLBI National Facility, Jodrell Bank Observatory, UK
\and
IPAC/Caltech, Mail Code 100-22, Pasadena, CA 91125, USA
\and
Department of Astronomy, AlbaNova 106 91 Stockholm, Sweden
\and
ESA/Space Telescope Science Institute, 
Baltimore, MD 21218, USA
\and
Harvard-Smithsonian Center for Astrophysics, 
Cambridge, MA 02138, USA
\and
National Radio Astronomy Observatory, P.O. Box 0, Socorro, NM 87801, USA
}
\maketitle

\begin{abstract}
The high resolution obtained through the use of VLBI gives 
an unique opportunity to directly observe the interaction
of an expanding radio supernova with its surrounding medium.
We present here results from our VLBI observations of 
the young supernovae SN~1979C, SN~1986J, and SN~2001gd.
\footnote{MAPT, AA, and ER thank the organizers for funding to attend the
conference.
This research has been supported by the Spanish DGICYT grants
AYA2001-2147-C02-01 and AYA2001-2147-CO2-02.
KWW wishes to thank the Office of Naval Research for the 6.1 funding
supporting this research.
}
\end{abstract}

\section{Introduction}
\label{sec:intro}

In the standard model of radio emission from supernovae, 
a blast wave is driven into an ionized,  
dense, slowly expanding wind. 
As a result, a high-energy-density shell is formed.
The relativistic electrons present in this shell spiral 
along the magnetic field and respond for the observed radio synchrotron emission.
The supernova quickly increases its radio brightness with time,
due to the increasingly smaller electron column density in the line
of sight. 
When the optical depth at cm-wavelengths is about unity, the supernova reaches
its maximum of emission,
after which the emission monotonically decreases due to expansion losses.
Very-Long-Baseline Interferometry (VLBI) observations of radio
supernovae are a powerful tool to probe the circumstellar interaction that takes
place after a supernova explodes. 
Indeed, high-resolution radio observations permit us to trace the presupernova mass
loss history by directly imaging the structure of the supernova as it
expands. The wealth of information includes also a
direct estimate of the deceleration of the supernova expansion,
estimates of the ejecta and circumstellar density profiles
(which has implications on the progenitor system), distortion of the
shock front, and the potential observation of Rayleigh-Taylor instabilities.
Unfortunately, the usually large distances to radio supernovae
makes their VLBI imaging a challenging task, and only radio supernovae
that are young, bright, nearby, and rapidly expanding are appropriate VLBI targets.
Since high-resolution radio observations of the young 
supernovae 1993J and 1987A, as well as VLBI observations of the supernova remnants in M82
are covered by other authors in these proceedings, we shall limit our
discussion to the cases of SN~1979C, SN~1986J, and SN~2001gd.

\section{SN~1979C}
\label{sec:sn1979c}
SN1979C was discovered on 19 April 1979 in the galaxy M100.
It was classified as a Type II Linear supernova, and had 
an expansion velocity of 9200 km/s, at an age of about 45 days~\cite{pan80}.
Its radio emission has been interpreted within the minishell model~\cite{chev82a} 
with some modifications~\cite{montes00}.
Recent studies of the environment of the supernova carried out in the
optical~\cite{vdyk99}  have put a possible constraint on the mass of
the progenitor of 17 to 18 \msun.
Previous VLBI observations of SN~1979C did not resolve the 
radio structure of the supernova ~\cite{bar85}. 
Nevertheless, the modeling of these radio data
showed the observations to be consistent
with an undecelerated expansion of the supernova 
($r \propto t^m, m=1$) for the first five years.

The radio light curves of SN~1979C are rather odd, 
showing a wiggling behavior for almost 20 yr. 
Recently, the radio
brightness of SN~1979C has stopped declining (or even started to
increase) and has apparently entered a new stage of
evolution~\cite{montes00}.
This new trend in the
radio light curves of SN~1979C is interpreted 
as being due to the supernova shock wave
having entered a denser region of material near the progenitor star.

\begin{figure}
\centering
\includegraphics[height=5cm, angle=0]{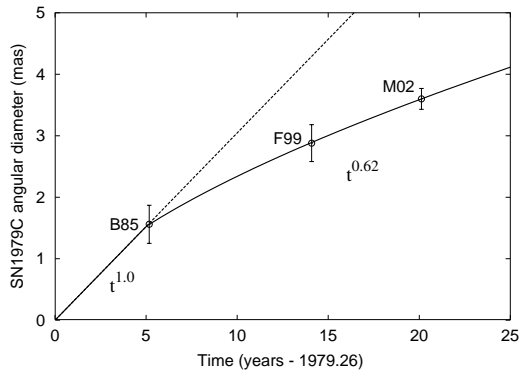}
\caption{
\scriptsize
Angular diameter of SN~1979C, in milliarcseconds, against time since
explosion. The solid line indicates a possible expansion, which goes
undecelerated ($m=1$) for the first five years, and strongly decelerated
($m=0.62$) from then on. See \cite{marca02} for details.
}
\label{fig:sn1979c}
\end{figure}

Prompted by this apparent new trend in the radio light curves of SN~1979C, 
we observed the supernova on June 1999 with a very sensitive 
four-antenna VLBI array at a wavelength of 18~cm.
Unfortunately, the VLBI array could not fully resolve the
radio structure of SN1979C, and we therefore
determined model-dependent sizes for the supernova and
compared them with previous results.
We estimated the size of the supernova by using three different models:
First, an optically thick, uniformly bright disk; 
second, an optically thin shell of width 30\% the outer shell radius;  
and third, an optically thin ring. 
(We refer the reader to~\cite{marca02}
for details of the modeling.)
The best-fit model was an optically thin shell model, 
which has an angular size of 1.80 milliarcseconds for the
outer shell supernova radius, corresponding to a linear size 
of $\sim 4.33 \EE{17}$ cm.
The combination of our VLBI observations ($t \sim 20$ yr, M02 in
Fig.~\ref{fig:sn1979c}) and previous ones~\cite{bar85} 
($t \sim 5$ yr, B85 in Fig.~\ref{fig:sn1979c}), plus optical observations 
~\cite{fesen99} at $t \sim 14$ yr (F99 in Fig.~\ref{fig:sn1979c}) 
allowed us to get an estimate of the epoch at which the deceleration started.
We estimate that the supernova shock was initially in free expansion
($m=1.0$)  for the first $6 \pm 2$ yr and then experienced a very strong
deceleration, characterized by a value of the deceleration parameter
of $m=0.62$.

If this deceleration is solely due to increased resistance from
the circumstellar medium (CSM), the mass of the CSM swept up by the
shock front, $M_{\rm swept}$, must be comparable to or larger than
the mass of ejected hydrogen-rich envelope, $M_{\rm env}$.
We estimate $M_{\rm swept}\,\sim\,1.6$\,\msun, assuming a
standard density profile for the CSM ($\rho_{wind} \propto r^{-2}$). 
Momentum conservation arguments then suggest that the mass of the hydrogen-rich
envelope ejected at explosion,  $M_{\rm env}$ is no larger than about
0.9\,\msun. 
Those results favor a binary star scenario for SN~1979C, as previously
suggested (e.g.~\cite{weiler86}). The low value of the hydrogen-rich envelope 
suggests that the companion of the
progenitor star stripped off most of the hydrogen-rich envelope mass
of the pre-supernova star prior to the explosion, similar to the
situation in SN~1993J.

Another point worth mentioning is the magnetic field in SN~1979C. 
If we assume equipartition between fields and particles, one expects a
minimum magnetic field in the range of 10--80\,mG to explain the 
observed level of radio emission.
Since the energy density of the wind magnetic field is not larger
than the kinetic energy density in the wind, it follows that
$B_w \lsim (\Mdot v_w)^{1/2}\,r^{-1}$ for a standard wind.
Using a mass-loss rate of $\sim 1.2 \EE{-4} \Msunyr$ for the
progenitor of SN~1979C \cite{lf88,weiler91} and a standard 
pre-supernova wind velocity $v_w$ of 10$\kms$, we obtain
$B_w \lsim\,0.2$\,mG, which is a factor of 50--400 lower than needed.
Therefore, if particles and fields are not very far from equipartition, 
then compression of the wind magnetic field due to the passage of the
supernova shock is not enough to explain the high magnetic fields 
in the supernova, and another
amplification mechanism must be acting. Turbulent amplification, as 
seems to be the case of SN1993J~\cite{torres01}, 
is likely the most promising mechanism.

\section{SN~1986J}
\label{sec:sn1986j}

The type II SN~1986J exploded in the galaxy NGC891
($D \approx 10$ Mpc). 
Unlike most supernovae, it was serendipitiously 
discovered in the radio more than three years after the explosion.
Modeling of the existing observations set the time of the explosion 
at the end of 1982 or the beginning of 1983~\cite{rupen87,weiler90}.
The supernova reached a peak luminosity at~$\wl 6$~cm of 
about 8 times larger than that of SN~1979C, and about 13 times larger than
the peak for SN~1993J, becoming one the brightest radio supernovae ever. 
Based upon its large radio luminosity,
the progenitor star was probably a red giant with
a main-sequence mass of $20-30 \Msun$ that had lost material
rapidly ($\Mdot \gsim 2 \EE{-4} \Msunyr$) in a dense stellar
wind~\cite{weiler90}.

VLBI observations made at $\wl 3.6$~cm
at the end of 1988 showed a shell-like structure for SN~1986J~\cite{bar91}.
The authors claimed the existence of several protrusions at
distances of twice the shell radius, and with apparent 
expansion velocities as high as $\sim 15000 \kms$.
Since these protrusions were twice as far as the mean radius of the
shell, it then follows that the main bulk of the shell expanded at roughly
$7500 \kms$.
Such protrusions have been successfully invoked~\cite{chugai99} to explain 
the coexistence of velocities smaller than 1000$\kms$ implied from the
observed narrow optical lines~\cite{rupen87,leibundgut91},
and the large velocities indicated from the VLBI measurements.

\begin{figure}
\centering
\includegraphics[width=12cm,angle=0]{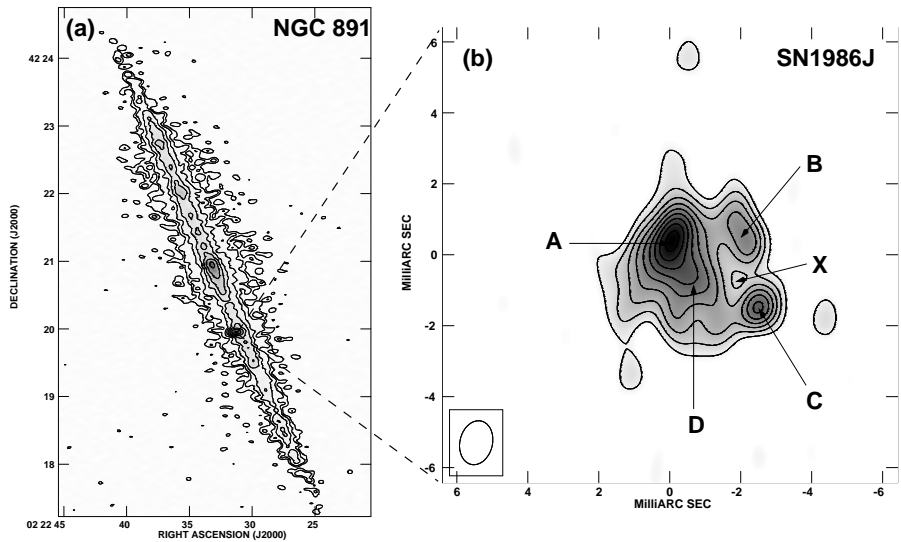}
\caption{
\scriptsize
VLA image (left) of the galaxy NGC~891 and its supernova SN~1986J, 
at the same frequency (5 GHz) and epoch of the global VLBI 
observations (right). See~\cite{torres01} for details.
}
\label{fig:sn1986j}
\end{figure}

We used archival VLA and global VLBI observations
of supernova 1986J at $\wl 6$~cm, taken about 16 yr after the
explosion, to obtain the images shown in Fig.~\ref{fig:sn1986j}
(see~\cite{torres02} for a comprehensive discussion).
The right panel corresponds to the $\wl 6$~cm VLBI image of SN1986J.
It shows a highly distorted shell of radio emission, indicative of a
strong deformation of the shock front.  
The apparent anisotropic brightness distribution 
is very suggestive of the forward shock
colliding with a clumpy, or filamentary wind. 
Note that there are several ``protrusions'' outside the shell,
though just above three times the noise level and at different position
angles from those previously reported~\cite{bar91}.
Therefore, these protrusions could not be real, but must be just artifacts of the image
reconstruction procedure. 
If this is the case, the disappearance
of the protrusions seen in the previous VLBI observations~\cite{bar91}
would imply a change in the density profile of the circumstellar wind. 

The angular size of the shell of SN~1986J on 21 February 1999 
is $\sim$4.7 mas, corresponding to a
linear size of 0.22 pc at the distance of SN~1986J.
Therefore, the average speed of the shell has decreased from around
$7500 \kms$ at the end of 1988 down to about $6300 \kms$ at the beginning
of 1999, indicating just a mild deceleration in the expansion of the
supernova ($m\approx\,0.90$).
We find a swept-up mass by the shock front of $\sim 2.2 \Msun$
for a standard wind density profile.
This large swept-up mass, coupled with the mild deceleration 
suffered by the supernova, suggests that the mass of the hydrogen-rich 
envelope ejected from the explosion was as large as $\sim 12 \Msun$.
This enormous value strongly indicates that the
supernova progenitor likely kept intact most of its
hydrogen-rich envelope by the time of explosion, and favours a
single, massive star progenitor scenario for SN~1986J.

We found a minimum total energy for the supernova (at the epoch of our
VLBI observations) in the range $(2-90) \EE{48}$ erg, depending on the ratio  
of the heavy particle energy to the electron energy. 
The corresponding values for the magnetic field should then
lie in the range $(13-90)$~mG, while the circumstellar wind 
magnetic field cannot exceed $\sim 0.3$~mG, i.e., it is 40 to 300 times
smaller than necessary to explain the observed radio emission.
As in the case of SN~1979C, 
turbulent amplification seems the most promising mechanism.

\section{SN~2001gd}
\label{sec:sn2001gd}

SN~2001gd was discovered 
on 24 November 2001 in the galaxy NGC 5033~\cite{itagaki01}, but
its exact explosion date is uncertain. 
A spectrum of SN~2001gd 
showed the supernova to be a type IIb event well past maximum light
~\cite{matheson01}.
The spectrum was almost identical to
one of SN~1993J obtained on day 93 after explosion~\cite{matheson00}.
The similarities of the optical spectra of SN~2001gd and SN1993J drove
us to observe SN~2001gd at the beginning of February 2002 with the VLA
~\cite{stockdale02}.
These and subsequent VLA observations confirmed the suggestion that 
SN~2001gd was a SN~1993J-like event, displaying a $\wl 6$~cm peak luminosity of 
about twice that of SN~1993J~\cite{stockdale03}.

\begin{figure}
\centering
\includegraphics[height=4cm]{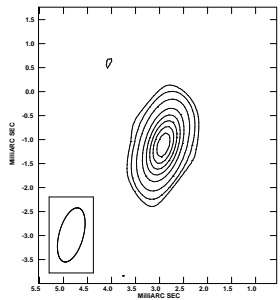}
\caption{
\scriptsize
Radio image of SN~2001gd obtained with GBT+Effelsberg+VLBA on 26 June
2002 
(total flux $\sim 3.9$~mJy; peak $\sim 3.3$~mJy beam$^{-1}$; 
rms noise $\sim 20 \muJ$;  beam $1.16 \times 0.50$ mas, 
P.A. $\sim -15^\circ$).
}
\label{fig:sn2001gd}
\end{figure}

We carried out high resolution VLBI observations
of SN~2001gd at $\wl 3.6$~cm, 
aimed at resolving the supernova structure.
Figure~\ref{fig:sn2001gd} shows the image of SN~2001gd obtained 
from our VLBI observations on June 2002. 
Since the VLBI observations did not resolve the radio
structure of SN~2001gd, we determined an angular size for the supernova
using model-dependent estimates.
We used the same three models as for SN~1979C.
However, none of the models was favoured at this stage and we only
inferred an upper limit on the expansion speed of the radio photosphere, 
which should lie in the range $\sim 23000 \kms$
to $\sim 26000 \kms$ at the epoch of our observations ($\sim 306$ days
after explosion) and for an assumed distance to the supernova of 21.6~Mpc.

We plan to follow the evolution of this  ``twin'' of SN~1993J
via VLBI observations of SN~2001gd, 
the first radio supernova for which such monitoring
is possible since the SN~1993J event in M~81.
Those future observations will likely allow us to
determine the deceleration and degree of self-similarity of the
supernova expansion, constrain the range of density profiles for the supernova progenitor
and the pre-supernova wind, and disentangle its radio emitting structure
as it interacts with the circumstellar medium.

We summarize in the table below the most remarkable results 
obtained from VLA and VLBI studies conducted on the radio supernovae we have
discussed here and, for comparison, those obtained on SN1993J (see
Alberdi \& Marcaide, this volume).

\begin{table}
\begin{center}
\begin{tabular}{lllll}
\hline\noalign{\smallskip}
                  &  SN~1979C     &  SN~1986J  &  SN~2001gd  & SN~1993J \\
\noalign{\smallskip}\hline\noalign{\smallskip}
Distance (Mpc)    &  $16.1\pm 1.3$         &  9.6  &  21.6? & $3.63\pm 0.34$     \\
Time since 
explosion$^1$(yr) &  $\sim 20.1$  &  $\sim 16$ &  $\lsim 1$    & $\sim 8.6$     \\
$(L/L_{\rm SN~1993J})_{6{\rm cm\,peak}}$    
                  &  $\sim 1.6$   &  $\sim 13$ &  $\sim 2$   &  1    \\
Optically thin phase?
                  &  Yes          &  Yes       & Likely yes  & Yes  \\
Radio brightness structure
                  &Shell (likely) &  Distorted shell & ?     & Smooth shell  \\
$\Mdot / 10^{-5} \Msunyr$  
                  &  $\sim (12-16)$ &  $\sim 20$ & ?           & $\sim 5$  \\
Deceleration parameter $(m)$   
                  &  $\sim 0.62$  &$\sim 0.90$ & ?           & $\sim 0.82$  \\
$t_{\rm break}$ (years) 
             &  $6 \pm 2$    & Not yet    & ?           & $\sim 0.5$ \\
Asymmetric expansion?    
                 &  No           & Yes        & ?           & No ($\lsim 5\%$) \\
Circumstellar medium
                 &  ?            & Clumpy     & ?           & Approx. smooth \\
$M_{\rm swept}/\Mdot$ 
           &  $\sim 1.6$   & $\sim 2.2$ & ?           & $\sim 0.4$ \\
$M_{\rm env}/\Mdot$ 
           &  $\sim 0.9$   & $\sim 12$  & ?           & $\sim 0.6$ \\
Explosion scenario 
           &  Binary       & Single     & ?           & Binary \\
Magnetic field amplification
           & Turbulent?     & Turbulent?  & ?           & Turbulent? \\
\noalign{\smallskip}\hline
\end{tabular}
\end{center}
\caption[]{\small 
$^1$Time since explosion when the VLBI observations were
carried out. The tabulated entries below refer to that time.
}
\label{tab:rsne}
\end{table}

\index{paragraph}
%
%

%
%

\printindex
\end{document}